\documentclass[aps,prl, preprint, superscriptaddress]{revtex4-1}

\usepackage{amsmath}
\usepackage[pdftex]{graphicx}
\usepackage[thinspace]{SIunits}

\newcommand{\bra}[1]{\left\langle #1\right|}
\newcommand{\ket}[1]{\left| #1\right\rangle}
\newcommand{\braket}[2]{\langle #1|#2\rangle}

\providecommand \BibitemShut  [1]{\csname bibitem#1\endcsname}%

\begin{document}

\preprint{}

\title{Non-linear two-photon resonance fluorescence on a single artificial atom}
\author{P.-L. Ardelt}
 \affiliation{Walter Schottky Institut and Physik-Department, Technische Universit\"at M\"unchen, Am Coulombwall 4, 85748 Garching, Germany \\}
\author{M. Koller}
 \affiliation{Walter Schottky Institut and Physik-Department, Technische Universit\"at M\"unchen, Am Coulombwall 4, 85748 Garching, Germany \\}
\author{T. Simmet}
 \affiliation{Walter Schottky Institut and Physik-Department, Technische Universit\"at M\"unchen, Am Coulombwall 4, 85748 Garching, Germany \\}
\author{L. Hanschke}
 \affiliation{Walter Schottky Institut and Physik-Department, Technische Universit\"at M\"unchen, Am Coulombwall 4, 85748 Garching, Germany \\}
\author{A. Bechtold}
 \affiliation{Walter Schottky Institut and Physik-Department, Technische Universit\"at M\"unchen, Am Coulombwall 4, 85748 Garching, Germany \\} 
\author{A. Regler}
 \affiliation{Walter Schottky Institut and Physik-Department, Technische Universit\"at M\"unchen, Am Coulombwall 4, 85748 Garching, Germany \\}
\author{J. Wierzbowski}
 \affiliation{Walter Schottky Institut and Physik-Department, Technische Universit\"at M\"unchen, Am Coulombwall 4, 85748 Garching, Germany \\}
\author{H. Riedl}
 \affiliation{Walter Schottky Institut and Physik-Department, Technische Universit\"at M\"unchen, Am Coulombwall 4, 85748 Garching, Germany \\}
\author{K. M\"uller}
\affiliation{Walter Schottky Institut and Physik-Department, Technische Universit\"at M\"unchen, Am Coulombwall 4, 85748 Garching, Germany \\}
 \affiliation{E. L. Ginzton Laboratory, Stanford University, Stanford, California 94305, USA\\}
\author{J.J. Finley}
 \affiliation{Walter Schottky Institut and Physik-Department, Technische Universit\"at M\"unchen, Am Coulombwall 4, 85748 Garching, Germany \\}
\email{finley@wsi.tum.de}

\date{\today}

\begin{abstract}
We report two-photon resonance fluorescence of an individual semiconductor artificial atom. By non-linearly driving a single semiconductor quantum dot via a two-photon transition, we probe the linewidth of the two-photon processes and show that, similar to their single-photon counterparts, they are close to being Fourier limited at low temperatures. The evolution of the population of excitonic states with the Rabi frequency exhibits a clear s-shaped behavior, indicative of the non-linear response via the two photon excitation process. We model the non-linear response using a 4-level atomic system representing the manifold of excitonic and biexcitonic states in the quantum dot and show that quantitative agreement is obtained only by including the interaction with LA-phonons in the solid state environment. Finally, we demonstrate the formation of dressed states emerging from a two-photon interaction between the artificial atom and the excitation field. The non-linear optical dressing induces a mixing of all four excitonic states that facilitates the tuning of the polarization selection rules of the artificial atom.   
\end{abstract}


\maketitle

Resonant interaction between light and matter forms the basis for much of quantum optics including dressing of atoms by the electromagnetic vacuum in high finesse cavities \cite{rempe1992} and the use of such cavity-QED techniques to interconvert between propagating and stationary quantum states \cite{Ritter2012}. It is exploited in the formation of microcavity polaritons, which optically established Bose-Einstein condensates \cite{kasprzak2006} with remarkable superfluidic properties \cite{Amo2009} and strong non-linearities suited for information processing in quantum optoelectronic devices \cite{Ballarini2013}. Resonance fluorescence emerging from the resonant interaction of a quasi two-level system with an incident coherent optical field possibly represents one of the simplest experiment that can be performed, but already leads to intriguing physics such as the appearance of the Mollow triplet \cite{Schuda1974, flagg2009, vamivakas2009, muller2007, xu2007coherent} and photon anti-bunching from a coherent driving field \cite{Kimble1977,Michler2000,Muller2014,matthiesen2013phase}. 

Over the past ten years, optically active quantum dots (QDs) have emerged as model solid-state systems to probe resonance fluorescence phenomena.  Their particular suitability for such experiments can be traced to their large oscillator strength and the two-level nature of their excitonic response \cite{schulte2015,gazzano2013, arnold2015}. Resonant driving of excitonic states via one-photon processes has facilitated single shot read-out of spins \cite{vamivakas2010, delteil2014} or phase-locking of indistinguishable single photons \cite{matthiesen2013phase}.  Moreover, resonance fluorescence techniques have led to the generation of ultra-coherent single photons from solid-state quantum emitters \cite{matthiesen2012, proux2015, nguyen2011} and, most recently, the observation of quadrature squeezed single photons \cite{schulte2015}. Whilst single photon resonance fluorescence has been extensively studied, resonant spectroscopy of \textit{two-photon} transitions has been limited to pulsed excitation schemes \cite{stufler2006, jayakumar2013} due their reduced oscillator strength \cite{stufler2006, jayakumar2013,Ardelt2014}. However, the coherent excitation of two-photon transitions is predicted to generate physical phenomena ranging from two-photon resonance fluorescence \cite{ sobolewska1977, mavroyannis1978,holm1985, alexanian2006, alexanian2007voigt} to two-mode squeezed states \cite{ficek1994} and photon bundles with exotic quantum statistics \cite{munoz2015}.

In this letter, we report two-photon resonance fluorescence studies on non-linear transitions of an individual optically active QD. We directly probe the linewidth of two-photon transitions and show it to be comparable to that of resonantly driven single photon transitions. The population evolution of the exciton and biexciton states exhibits clear signatures of non-linear light-matter interaction in excellent agreement with an atomic 4-level system including the coupling to LA - phonons. Finally, we directly observe dressed states in the non-linearly driven system and demonstrate how the polarization selection rules can be tuned by dressing.  

The sample investigated was grown using molecular beam epitaxy. On top of a 14-pair $\lambda / 4$ GaAs/AlAs distributed Bragg reflector (DBR) a layer of low density InGaAs quantum dots (QDs) is centred in a nominally $260 \, \nano\meter$ thick GaAs layer acting as a weak microcavity to enhance the photon extraction efficiency from the surface of the sample \cite{gao2012}. An n-doped layer below the QDs and a Ti/Au metal top contact form a Schottky diode that facilitates control of the electric field and the energies of excitonic transitions of the QD using the DC Stark shift \cite{warburton2000}.

\begin{figure}[t]
\includegraphics[width=1\columnwidth]{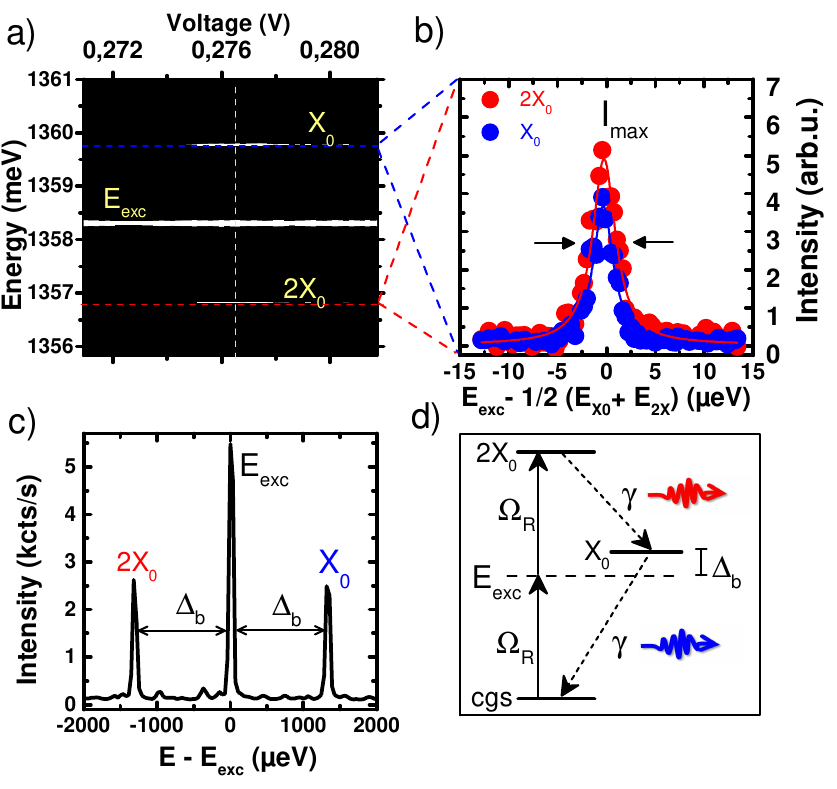}
\caption{\label{fig:Figure1} (Color online) (a) Voltage dependent luminescence of the neutral exciton $X_{0}$ and biexciton $2X_{0}$ transitions tuning the excitation laser $E_{exc}$ across the two-photon resonance $cgs \rightarrow \rightarrow 2X_{0}$. (b) Detuning-dependent luminescence intensities $I_{0}$ and $I_{2X_{0}}$ to characterize the two-photon transition $cgs \rightarrow \rightarrow 2X_{0}$. (c) Spectrum excited on resonance. (d) Energy level structure of a QD driven on the two-photon resonance with a Rabi energy $\Omega_{R}$.}
\label{Figure1}
\end{figure}

To characterize the two-photon excitation of individual QDs, we operate the Schottky diode on the neutral charge stability plateau at a lattice temperature of $T = 4.2 \, \kelvin$ \cite{Ardelt2014}. The energy level structure of the QD is presented in Fig. \ref{Figure1}d. While $cgs$ represents the crystal ground state of the system, the first optical excitation corresponds to a charge neutral exciton $X_{0}$ (a single excited electron-hole pair $e-h$) and the second optical excitation to a charge neutral biexciton $2X_{0}$ (two $e-h$ pairs). The optically active one-photon transitions from $2X_{0} \rightarrow X_{0}$ and $X_{0} \rightarrow cgs$ are detuned by $2 \Delta_{b}$ with respect to each other due to the different attractive Coulomb interactions in the $X_{0}$ and $2X_{0}$ states respectively \cite{warburton2000, Ardelt2014}. This detuning $2\Delta_{b}$ allows to directly address the biexciton state $2X_{0}$ from the $cgs$ via a resonant two-photon excitation process $cgs\rightarrow\rightarrow E_{2X_{0}}$ by red detuning the excitation laser from the $X_{0} \rightarrow cgs$ transition by half the binding energy $\Delta_{b} = \frac{1}{2}(E_{X_{0}} - E_{2X_{0}})$. Here, the double arrow denotes the two-photon nature of the transition.

In Fig. \ref{Figure1}a, we fix the energy of the excitation laser $E_{exc} = \hbar \omega_{L}$ close to the two-photon resonance (red detuned from $X_{0}$ by $\sim \Delta_{b}$) and fine-tune the energy of $2X_{0}$ relative to it using the DC Stark effect \cite{warburton1998} by applying a DC-voltage $V$ across the Schottky diode. At a bias of $V = 0.2762 \, \volt$ we clearly resolve emission from the $2X_{0}$ and $X_{0}$ transitions in Fig. \ref{Figure1}c signifying that the two-photon resonance condition has been met. Here, the excitation laser is positioned exactly on resonance at $E_{exc}=\frac{1}{2}(E_{X_{0}}+ E_{2X_{0}})$. The simultaneous emission of fluorescence results from the cascaded recombination of $2X_{0}$ via the two \emph{single-photon} transitions $2X_{0} \rightarrow X_{0} \rightarrow cgs$ \cite{Ardelt2014} indicated in Fig. \ref{Figure1}d and is a clear signature of resonant two-photon excitation of the biexciton $cgs \rightarrow\rightarrow 2X$ \cite{stufler2006, jayakumar2013, Ardelt2014}.  

As we drive the two-photon transition $cgs \rightarrow\rightarrow 2X_{0}$ using a cw laser with a narrow bandwidth of $\sim 50 \, \nano \electronvolt$, measurements of the absorption linewidth of the two-photon transition are facilitated. In Fig. \ref{Figure1}b, we present the integrated intensity of the single-photon transition $2X_{0} \rightarrow X_{0}$ in red ($X_{0} \rightarrow cgs$ in blue) as a function of the excitation detuning from the two-photon resonance. \footnote{The excitation laser detuning is calculated from Fig. \ref{Figure1}a using the DC Stark effect of the biexciton state $2X_{0}$ \cite{warburton1998}. Thus, the plotted intensities in Fig. \ref{Figure1}b directly reflect the absorption spectrum of the two-photon transition $cgs \rightarrow\rightarrow 2X$ that displays a clear Lorentzian line-shape as indicated by the fits.} We performed this basic characterization on five different QDs all of which produced results similar to those in Fig. \ref{Figure1}b. For excitation powers of $P_{exc}= 15 \, \micro \watt$ we obtain linewidths as small as $\Delta \omega_{ 2X_{0} \rightarrow X_{0} } = 2.19 \, \micro \electronvolt$ ($\Delta \omega_{X_{0} \rightarrow cgs} = 2.14 \, \micro \electronvolt$) similar to the minimal absorption linewidth of the single photon transitions of the same QD $\Delta \omega_{cgs \rightarrow X_{0}} = 1.9 \, \micro \electronvolt$. In agreement with the narrow linewidths, Michelson interferometry performed on the emitted photons reveals photonic coherence times up to $T_{2} (2X_{0}) = 432 \, \pico \second$ ($T_{2}(X_{0}) = 414 \, \pico \second $) at low excitation powers (see supplementary). This indicates near Fourier limited coherences within the driven 4-level QD system.   

\begin{figure}[t]
\includegraphics[width=1\columnwidth]{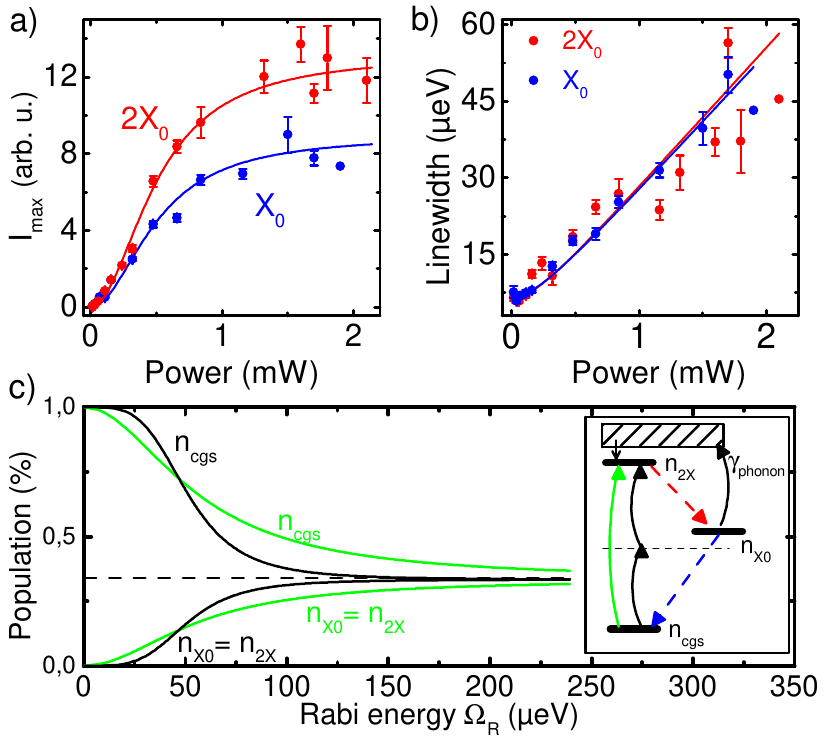}
\caption{\label{fig:Figure2} (Color online) (a) S-shaped power dependence of the maximum intensities $I_{max}$ with theoretically calculated intensities from the population distribution $n_{i}$ including LA-phonon coupling (schematically presented on the inset in (c)). (b) Linewidth of the two-photon transition as function of the excitation power. (c) Calculated evolution of the populations $n_{i}$ under resonant two-photon excitation (black) and single-photon excitation (green) \emph{without} LA-phonon coupling.}
\label{Figure2}
\end{figure}

To develop a quantitative understanding we model the QD as a four level system consisting of the states $\ket{cgs}$, $\ket{X_{0}(V)}$,$\ket{X_{0}(H)}$, $\ket{2X_{0}}$ driven by a semi-classical electromagnetic field with Rabi energy $\Omega_{R}$. Modeling was performed using the Quantum Toolbox in Python \cite{qutip, qutip2}. Here, the single exciton state $X_{0}$ in Fig. \ref{Figure1}d is replaced by the two states $X_{0}(V)$ and $X_{0}(H)$ that emerge from the anisotropic $e-h$ exchange interaction and are split by $\delta_{0}$. We consider the excitation field to be on resonance with the two-photon transition $\hbar \omega_{L} =  \frac{1}{2}(E_{X_{0}} + E_{2X_{0}})$, change into a frame rotating with the laser frequency $\omega_{L}$ and use the rotating wave approximation. The Hamiltonian in matrix form $H_{ij} = \bra{j} \hat{H} \ket{i}$ in the bare state basis $i,j = cgs$, $X_{0}(V)$, $X_{0}(H)$, $2X_{0}$ then reads:

\begin{equation*}
\label{eq:matrix}
H_{total} = H_{0} + H_{int} = \begin{pmatrix}
	0 & -\frac{\Omega_{V}}{2} & -\frac{\Omega_{H}}{2} & 0  \\ 
	-\frac{\Omega_{V}}{2} & \Delta_{b} - \frac{\delta_{0}}{2} & 0 & -\frac{\Omega_{V}}{2} \\
	-\frac{\Omega_{H}}{2} & 0 &  \Delta_{b} + \frac{\delta_{0}}{2} & -\frac{\Omega_{H}}{2}\\
	0 & -\frac{\Omega_{V}}{2} &  -\frac{\Omega_{H}}{2} & 0 \\
 \end{pmatrix}
\end{equation*}

Note, we assign similar Rabi energies $\Omega_{H(V)}=\Omega_{R}$ to both single photon transitions corresponding to a diagonally polarized excitation field $\ket{D} = \frac{1}{\sqrt{2}} (\ket{H} + \ket{V})$ as present in our cross-polarized resonance fluorescence set up. We only allow the optically active transitions $cgs \rightarrow X_{0}$ and $X_{0} \rightarrow 2X$ and include the spontaneous emission between $2X \rightarrow X_{0}(H/V)$ and $X_{0}(H/V) \rightarrow cgs$ as collapse operators with rates $\gamma_{i}$ as illustrated in Fig. \ref{Figure1}d (details supplementary).   

The excitation laser couples $\ket{cgs}$ and $\ket{2X_{0}}$ via two-photon excitation. Thus, for the steady state solution of the system $\rho_{ss}$, the coupling $cgs \rightarrow \rightarrow 2X_{0}$ leads to a redistribution of the populations $n_{i} = Tr(\rho_{ss} \ket{i}\bra{i})$ between the four states $i = cgs, X_{0}(V), X_{0}(H), 2X_{0}$. In Fig. \ref{Figure2}a, we present the maximum intensities $I_{max}$ of the two-photon absorption spectra (similar to Fig. \ref{Figure1}b) as a function of increasing excitation power $P_{exc}$ for both transitions $2X_{0} \rightarrow X_{0}$ and $X_{0} \rightarrow cgs$. The relative intensities $I_{max}$ are directly proportional to the populations of the upper state. Thus, the biexciton population is proportional to $I_{max}(2X_{0}) \propto n_{2X_{0}}$. As we do not resolve the fine structure splitting $\delta_{0}$ between $X_{0}(V)$ and $X_{0}(H)$ due to the limited resolution of the CCD-detection, the exciton population is proportional to the sum of the populations $I_{max}(X_{0}) \propto (n_{X_{0}(V)} + n_{X_{0}(H)})$. 

We clearly resolve two features in Fig. \ref{Figure2}a : First, an s-shaped power dependence of the population evolution and, second, an increased population redistribution into $2X_{0}$ compared to $X_{0}$ for increasing $P_{exc}$. The s-shape is qualitatively understood by considering the non-linear nature of the excitation process: For a resonantly driven two-photon transition, its Rabi energy depends on the square of the single-photon Rabi energy $\Omega_{R}$ with $\Omega \propto \frac{\Omega_{R}^{2}}{\Delta}$ \cite{linskens1996} and induces an additional curvature in the power dependent saturation of the populations $n_{i}$. We present the numerically calculated evolution of the steady state populations $n_{i}$ for the states $cgs$, $X_{0}$ and $2X_{0}$ in Fig. \ref{Figure2}c. All populations $n_{i}$ saturate against a relative value of $1/3$ with increasing $\Omega_{R}$. For comparison, the green curve in Fig. \ref{Figure2}c shows the evolution of the populations $n_{i}$ for the same system excited via a single-photon transition from a semi-classical light field in resonance with the $cgs \rightarrow 2X_{0}$ transition. We clearly observe, that the saturation behavior of the populations $n_{i}$ towards $1/3$ remains, however the s-shape of the power dependency is flattened.

While the shape of the population evolution is well reproduced by the model, the increasing distribution of population into the biexciton state $n_{2X} > n_{X_{0}}$, observed in Fig. \ref{Figure2}a, is not. In order to model the redistribution we extend the Hamiltonian $ H_{QD} + H_{int}$ by including coupling to the longitudinal-acoustic (LA) phonon bath in the GaAs environment:

\begin{equation}
H_{QD - phonon}= \sum_{i} \ket{i} \bra{i} \sum_{q} \lambda_{q}^{i}  (\hat{b_q}^\dagger+\hat{b_q} )
\end{equation}

where $\hat{b_q}^\dagger$ ($\hat{b_q}$) describes the creation (annihilation) operator of an LA-phonon with a wave vector $q$ and energy $\hbar\omega_q$. The detuning dependent coupling strength between the charge carrier states $i={X_0(V), X_0(H), 2X_{0}}$ and the LA-phonon reservoir is described by $\lambda_{q}^{i}$. Most importantly, the resulting charge carrier - phonon interaction spectrum $J(\Delta \omega)$ is highly sensitive to the energy detuning of the excitation laser $\hbar \omega_{L}$. Only for positive detunings $\Delta \omega_{i \rightarrow j} = \omega_{L} - \omega_{i \rightarrow j} > 0$ of the excitation laser with respect to the optical transition $i \rightarrow j$, the spectrum acquires considerable values since LA - phonons are frozen out at low temperatures and phonon emission processes dominate \cite{quilter2015, weiler2012, bounouar2015, glassl2013}. When the excitation laser resonantly drives $cgs \rightarrow \rightarrow 2X_{0}$, it is blue detuned from $2X_{0} \rightarrow X_{0}$ ($\Delta \omega_{2X \rightarrow X_{0}} = +\Delta_{b} > 0$) and red detuned from $X_{0} \rightarrow cgs$ ($\Delta \omega_{2X \rightarrow X_{0}} = -\Delta_{b} < 0$). Thus, although we include both interactions in the Hamiltonian $H_{QD-phonon}$, it influences only the population redistribution between $X_{0}$ and $2X_{0}$.

For a blue detuning of the excitation laser of $\Delta \omega_{2X_{0} \rightarrow X_{0}} = \Delta_{b} = +0.93 \, \milli \electronvolt$, the charge carrier LA-phonon interaction will lead to an incoherent population transfer from $X_{0}$ to $2X_{0}$ via a phonon assisted excitation process \cite{weiler2012} (inset on Fig.\ref{Figure2}c): It works effectively as an incoherent population pump-rate $\gamma_{phonon}$ from $X_{0}$ to $2X_{0}$. This pump rate $\gamma_{phonon}$ increases the steady state population of the biexciton $n_{2X_{0}}$ with respect to the exciton $n_{X_{0}}$. To calculate the resulting population redistribution, we transform the Hamiltonian $H=H_{QD} + H_{int}+ H_{QD - phonon}$ into the polaron frame (details supplementary) \cite{ge2013}. The resulting relative populations are fit to the intensities of the single photon transitions $I_{2X \rightarrow X_{0}} \propto n_{2X_{0}}$ and $I_{X_{0} \rightarrow cgs} \propto n_{X_{0}}$ in Fig. \ref{Figure2}a by varying the dipole moment of the two-photon transition. We achieve excellent agreement between the calculated populations $n_{2X_{0}}$ and $n_ {X_{0}}$ and the recorded intensities by including the charge carrier - LA phonon interaction. 

We remark, that from the saturating behavior of the emission intensity with increasing $\Omega_{R}$, we expect the linewidth of the two-photon transition $\Delta \omega_{cgs \rightarrow \rightarrow 2X}$ to exhibit power broadening. Indeed, plotting the linewidth of two-photon absorption spectra in Fig. \ref{Figure2}b for various $P_{exc}$, we observe a clear increase of $\Delta \omega_{cgs \rightarrow\rightarrow 2X_{0}}$ with increasing Rabi energy $\Omega_{R}$. Although the linewidth broadening can be fitted using a two-photon power broadening, we emphasize that this is unlikely to be the only source of linewidth broadening. For increasing $P_{exc}$ photon mediated charging events may lead to charge noise in the QD environment \cite{nguyen2012optically, houel2012}. Note, that we expect spin noise to make a minor contribution on the linewidth broadening of $cgs \rightarrow\rightarrow 2X_{0}$ since both the initial and final states $cgs$ and $2X_{0}$ have a total spin projection of $S=0$ \cite{kuhlmann2013noise}. 

\begin{figure}[t]
\includegraphics[width=1\columnwidth]{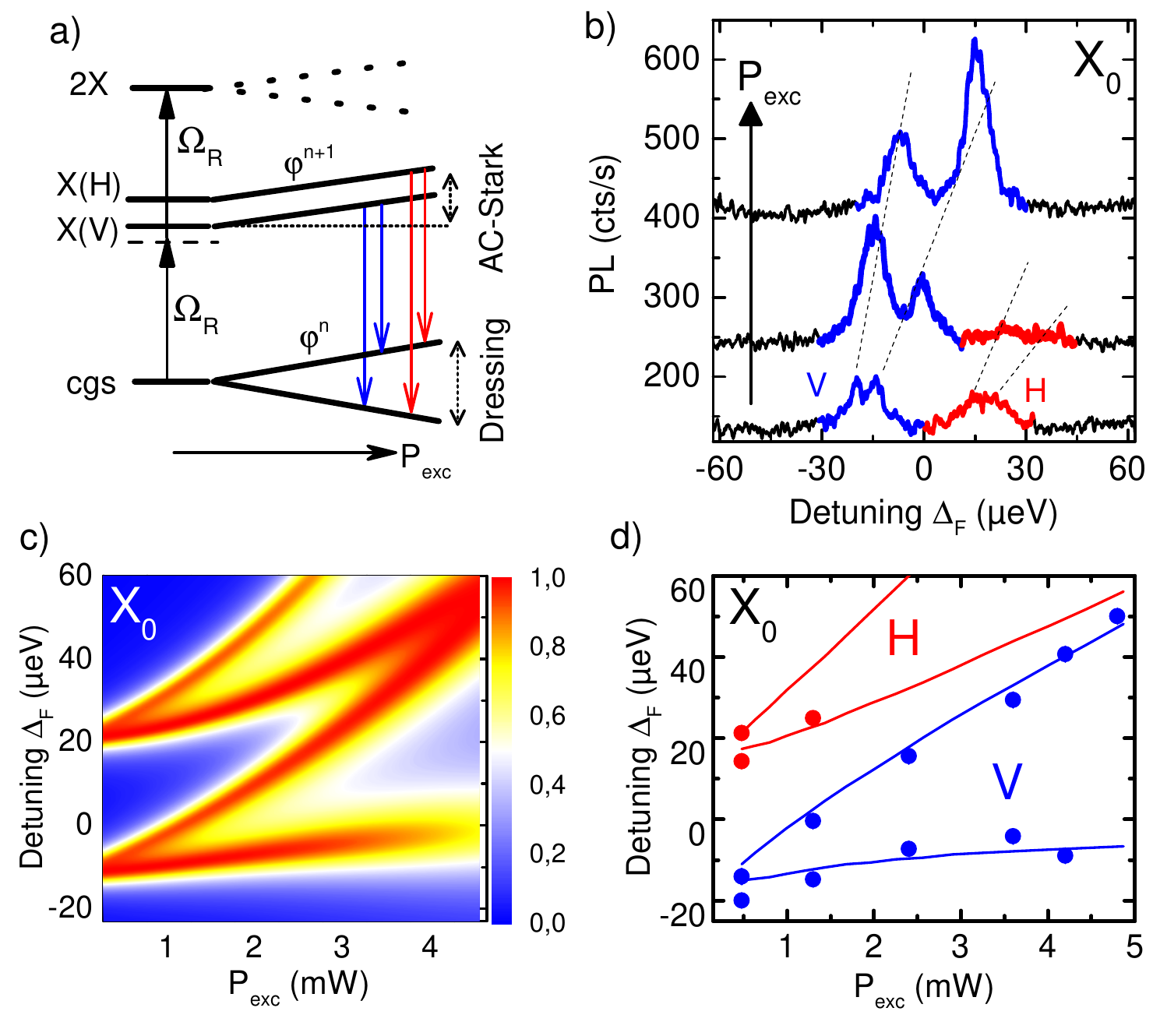}
\caption{\label{fig:Figure4} (Color online) (a) Schematic evolution of the states $cgs$, $X_{0}(H/V)$, $2X_{0}$ under resonant two-photon excitation with Rabi frequency $\Omega_{R}$ leading to the formation of dressed eigenstates $\varphi_{i}$. (b) Sample Fabry-Perot spectra of the $X_{0} \rightarrow cgs$ transitions for increasing excitation power $P_{exc}$. (c) Calculated optical transitions evolving from bare state $X_{0}(H(V)) \rightarrow cgs$ to dressed state $\varphi_{j}^{n+1} \rightarrow \varphi_{i}^{n}$ transitions as a function of $P_{exc}$. (d) Peak positions from experiment and theory in (b) and (c), respectively.}
\label{Figure4}
\end{figure}

Finally, we discuss the emergence of new dressed eigenstates of the LA - phonon reservoir coupled four-level system under resonant two-photon excitation $cgs \rightarrow \rightarrow 2X_{0}$. We consider two effects: First, the driving field with Rabi frequency $\Omega_{R}$ couples $\ket{cgs}$ to $\ket{2X_{0}}$ to form new non-linearly dressed eigenstates, illustrated in Fig. \ref{Figure4}a as a splitting of the bare states $E_{2X_{0}}$ and $E_{cgs}$. Second, the excitation field $\hbar \omega_{L}$ is detuned from the single photon transitions $cgs \rightarrow X_{0}(H/V)$ and $X_{0}(H/V) \rightarrow 2X_{0}$. However, for high $P_{exc}$ the states of the single-photon transitions are coupled via the AC-Stark effect $\Delta_{AC}$ as indicated by a shift in Fig. \ref{Figure4}a. Thus, the Rabi frequency $\Omega_{R}$ will admix all four bare states into new dressed eigenstates $\varphi_{1-4}$. In the experiment we remain in a regime, where for the highest excitation power the AC Stark shift is comparable to the exchange energy $\Delta_{AC} \sim \delta_{0}$ and the Rabi frequencies does not exceed the binding energy $\Delta_{b}$. Thus, the fine structure $\delta_{0}$ and binding energy $\Delta_{b}$ do not prevent the formation of dressed states $\ket{\varphi_{1-4}} = \sum_{i} \alpha_{i}^{1-4} \ket{i}$ but lead to unequal, power dependent coefficients $\alpha_{i}$ of the different bare state components $i = cgs, X_{0}(V), X_{0}(H), 2X_{0}$ (details supplementary).

To experimentally resolve the non-linear dressing of the states, we performed high resolution spectroscopy on the $X_{0}(V/H) \rightarrow cgs$ transitions using a Fabry Perot interferometer with a free spectral range of $FSR = 124 \, \micro \electronvolt$ and a resolution finer than $< 1 \, \micro \electronvolt$. Exemplary spectra are presented in Fig. \ref{Figure4}b. For the lowest $P_{exc}$, we already resolve that each fine structure peak of the transitions $X_{0}(V/H) \rightarrow cgs$ is split into two peaks due to the non-linear dressing of the system. For increasing $P_{exc}$, the splitting emerging from $X_{0}(V) \rightarrow cgs$ is increased. At the same time the second set of transitions from $X_{0}(H) \rightarrow cgs$ is strongly suppressed. These observations are entirely in accord with our expectations and result from a modification of the polarization selection rules of the optical transitions upon dressing. Increasing the driving $\Omega_{R}$ of the diagonally $D$ polarized excitation laser, the dressed eigenstates $\varphi_{2-4}$ evolve into a coherent superposition of the bare eigenstates $cgs$, $X_{0}(V)$, $X_{0}(H)$ and $2X_{0}$. As transitions from and to the bare states $X_{0}(H/V)$ are addressed by the linear polarization $H/V$, their superpositions are addressed by superpositions of $H$ and $V$ \cite{muller2013all}. Accordingly, the polarization of photons from transitions between the dressed states $\varphi_{j}^{n+1} \rightarrow \varphi_{i}^{n}$ (of the manifolds $n+1$ and $n$) is \textit{rotated} with respect to the polarization $H/V$. In our experimental set up, we use a $\sim D/A$ cross-polarized excitation/detection scheme to suppress stray light from the excitation laser. Thus, we only resolve the two transitions emerging from $X_{0}(V) \rightarrow cgs$ in Fig. \ref{Figure4}b that are polarized predominantly along $\sim A$, while the second set of transitions polarized $\sim D$ is suppressed (details supplementary). 

To obtain quantitative results we numerically calculate the evolution of the dressed state transitions $\varphi_{i}^{n+1} \rightarrow \varphi_{j}^{n}$ emerging from $X_{0}(H/V) \rightarrow cgs$. Fig. \ref{Figure4}c presents the spectrum from the Hamiltonian $H = H_{QD} + H_{Int} + H_{QD - phonon}$ that also includes the LA-phonon interaction. The initial transitions between $X_{0}(V) \rightarrow cgs$ ($X_{0}(H) \rightarrow cgs$) are split with increasing excitation power and shifted to higher energies as discussed above. The linewidth of the transitions exhibits an additional broadening due to the charge carrier - LA phonon coupling. Finally, we plot in Fig. \ref{Figure4}d the extracted peak positions from the high resolution fluorescence spectra in Fig. \ref{Figure4}b together with the calculated energies of the dressed state transitions in Fig. \ref{Figure4}c and observe excellent agreement. Importantly, no further fitting parameters were included after previously adapting the model to the experimentally measured population distributions (Fig. \ref{Figure2}a). This corresponds to an evolution into dressed states of the non-linearly driven system, the hallmark of two-photon resonance fluorescence.     

In summary, we presented two-photon resonance fluorescence studies on a single QD. Monitoring the populations of excitonic states, we found clear signatures of non-linear light-matter interactions and coupling to LA - phonons. Finally, we demonstrated the formation of non-linearly dressed states due to the two-photon excitation of the QD. Our results pave the way for investigating a wealth of optical phenomena resulting from the non-linear light matter interaction ranging from squeezed light \cite{ficek1994, huang2014} to the generation of photons with arbitrary quantum statistics \cite{munoz2015}.

We gratefully acknowledge financial support from the DFG via SFB-631, Nanosystems Initiative Munich, the EU via the ITN S\textsuperscript{3} Nano and BaCaTeC. KM acknowledges support from the Alexander von Humboldt foundation and the ARO (grant W911NF-13-1-0309).

\newpage
\section{Supplementary: Non-linear two-photon resonance fluorescence on a single artificial atom}

The supplementary is organized as follows: In the first section, we will present measurements of the photonic coherences under resonant two-photon excitation. In the second section, we analyze the composition of the new dressed eigenstates of the Hamiltonian $H_{QD} + H_{int}$ and analyze the polarization of the transitions between them. In the third section, we present full details of the theoretical model used in the letter. In the fourth section, we present additional measurements of the incoherent phonon assisted excitation process that further support the conclusions drawn in the main manuscript.   

\section{Photonic coherences under resonant two photon excitation}
\label{sec:coherences}

\begin{figure}
\includegraphics[width=1\columnwidth]{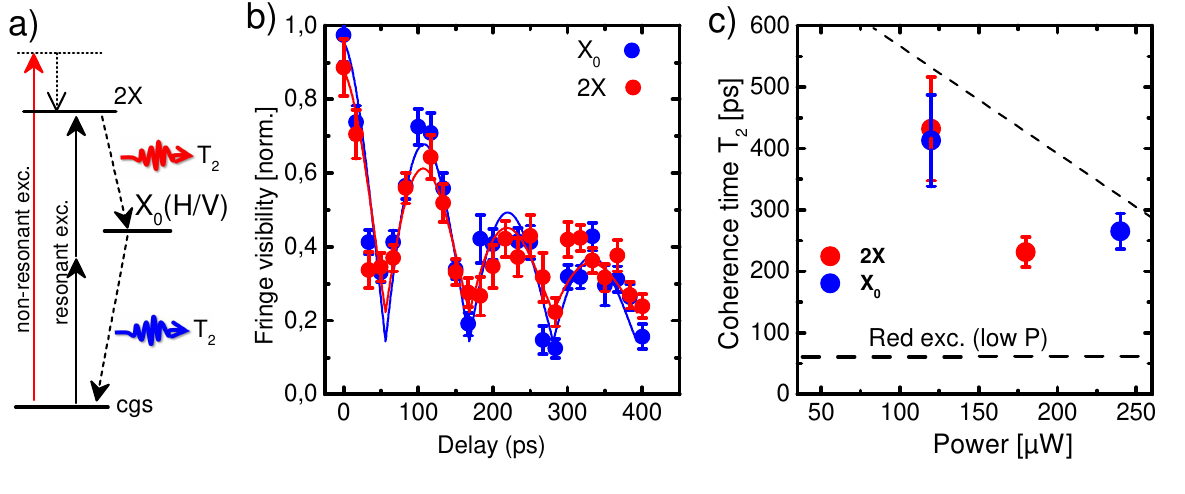}
\caption{\label{fig:Figure1S} (Color online) (a) Excitation scheme for two-photon resonant excitation (black) and above band gap excitation (red). We record the photonic coherences $T_2$ generated by both transitions $X_{0} \rightarrow cgs$ and $2X_{0} \rightarrow X_{0}$  (b) Fringe visibility as a function of time delay $\tau$ in Michelson interferometry. Both transitions $2X_{0}$ and $X_{0}$ exhibit a beating due to the fine structure split of the neutral exciton state $X_{0}$. (c) Photonic coherence times $T_{2}$ extracted from the decay of the Michelson interferometry experiments in (b) for resonant two-photon excitation at different powers.}
\label{Figure1S}
\end{figure}

To determine the quality of the photon coherences quantitatively, we perform Michelson interferometry on the photons generated by resonant two-photon excitation of the QD biexciton state $2X_{0}$ as schematically illustrated in Fig.\ref{Figure1S}a. We resonantly generate population $n_{2X_{0}}$ in the biexciton state that decays via the cascade $2X \rightarrow X_{0} \rightarrow cgs$ emitting a blue and red detuned photon with respect to the excitation laser energy $E_{exc}$ (black). In figure \ref{Figure1S}b we present the fringe visibilities for photons from both transitions, $2X_{0} \rightarrow X_{0}$ and $X_{0} \rightarrow cgs$, for resonant two-photon excitation of the transition $cgs \rightarrow \rightarrow 2X_{0}$ with a power of $P_{exc} = \, 120 \micro \watt$. The resulting photon coherence times $T_{2}$ from fits of the fringe visibility decay in Fig.\ref{fig:Figure1S}b are presented in Fig. \ref{fig:Figure1S}c for different excitation powers. Note, that we do not observe a single exponential decay of the fringe visibility in Fig.\ref{fig:Figure1S}b due to the anisotropic exchange interaction that splits the neutral exciton $X_{0}$ into two states $X_{0}(V)$ and $X_{0}(H)$ \cite{bayer2002fine}. The fine structure splitting induces a beating with a frequency corresponding to the frequency of exchange interaction. We use an exponential decay multiplied with a cosine function to include the fine structure beating in the decay of the visibility to extract the coherence times $T_{2}$.

As depicted in Fig. \ref{Figure1S}c, we observe photon coherences for the $2X_{0} \rightarrow X_{0}$ and $X_{0} \rightarrow cgs$ transitions up to $T_{2}= 432 \, \pico\second$ and $T_{2} = 414 \, \pico \second$, respectively. For increasing excitation powers $P_{exc}$, the photonic coherences are reduced. However, as indicated in figure \ref{fig:Figure1S}c, the photon coherences exceed the coherence times for non-resonant excitation above the band gap with a red laser diode $E_{exc} = 1952.5 \, \milli \electronvolt$. For different QDs the corresponding coherence times did not exceed $T_{2} < 100 \, \pico \second$ for excitation close to the QD neutral exciton transition. We carefully suggest that the improved photonic coherences originate from reduced charge carrier fluctuations in the QD environment due to the resonant two photon excitation scheme \cite{kiraz2004, nguyen2012optically, Muller2014}. For non-resonant excitation with moderate powers, additional QDs are charged in the high density QD-layer. Furthermore, the above band gap excitation will lead to charging events of charge traps in the QD environment \cite{nguyen2012optically, houel2012}. The associated charge noise reduces the photonic coherence times $T_{2}$ and limits the potential for applications of the non-resonant excitation scheme compared to resonant two-photon excitation for e.g. using the QD-biexciton cascade $2X_{0} \rightarrow X_{0}(H/V) \rightarrow cgs$ as a source of entangled photons \cite{Muller2014}.

\section{Analysis of dressed states and polarizations of the dressed state transitions}
\label{sec:dressed}

In this section we provide an intuitive picture of the 4-level atomic system that is non-linearly driven by resonant two-photon excitation and discuss the polarizations for transitions between the emerging dressed states. We consider the Hamiltonian $H = H_{QD} + H_{int}$ in a frame rotating with the laser frequency $\omega_{L}$. The Hamiltonian in matrix form $H_{ij} = \bra{j} \hat{H} \ket{i}$ using the rotating wave approximation and the bare state basis $i,j = cgs$, $X_{0}(V)$, $X_{0}(H)$, $2X_{0}$ then reads (as in the main section of the letter):

\begin{equation*}
\label{eq:matrix}
H_{total} = H_{0} + H_{int} = \begin{pmatrix}
	0 & -\frac{\Omega_{V}}{2} & -\frac{\Omega_{H}}{2} & 0  \\ 
	-\frac{\Omega_{V}}{2} & \Delta_{b} - \frac{\delta_{0}}{2} & 0 & -\frac{\Omega_{V}}{2} \\
	-\frac{\Omega_{H}}{2} & 0 &  \Delta_{b} + \frac{\delta_{0}}{2} & -\frac{\Omega_{H}}{2}\\
	0 & -\frac{\Omega_{V}}{2} &  -\frac{\Omega_{H}}{2} & 0 \\
 \end{pmatrix}
\end{equation*}

Here, $\Delta_{b}$ denotes the half the binding energy $\frac{E_{B}}{2}$ and $\delta_{0}$ the fine structure of the neutral exciton $X_{0}$ with the values presented in table \ref{tab:parameters} in the next section. We consider excitation with a diagonally polarized $\ket{D} = \frac{1}{\sqrt{2}} (\ket{H} + \ket{V})$ semi-classical electro-magnetical field such that $\Omega_{V} = \Omega_{H} = \Omega_{R}$. In the high Rabi frequency limit $\Delta_{b}, \delta_{0} << \Omega_{r} \rightarrow \infty$ where the Rabi energy $\Omega_{R}$ is the only relevant energy scale, the new eigenstates of the 4-level QD system will converge towards: (i) a zero energy state $\ket{\varphi_{1}} = \frac{1}{\sqrt{2}} (\ket{cgs} + \ket{2X_{0}})$, an admixed exciton state $\ket{\varphi_{2}} = \frac{1}{\sqrt{2}} (\ket{X_{0}(H)} - \ket{X_{0}(V)})$ and two states that are a superposition of all four bare states (iii) $\ket{\varphi_{3}} = \frac{1}{2} (\ket{cgs} + \ket{X_{0}(H)} + \ket{X_{0}(V)} + \ket{2X_{0}})$ and (iv) $\ket{\varphi_{4}} = \frac{1}{2} (- \ket{cgs} + \ket{X_{0}(H)} + \ket{X_{0}(V)} - \ket{2X_{0}})$. However, as we operate the QD in a regime where $\Omega_{R} < \Delta_{b}$ the binding energy (and to a certain extent the fine structure $\delta_{0}$) prevents the states $\varphi_{1/2/3/4}$ from achieving a perfectly equal admixture. 

\begin{figure}
\includegraphics[width=1\columnwidth]{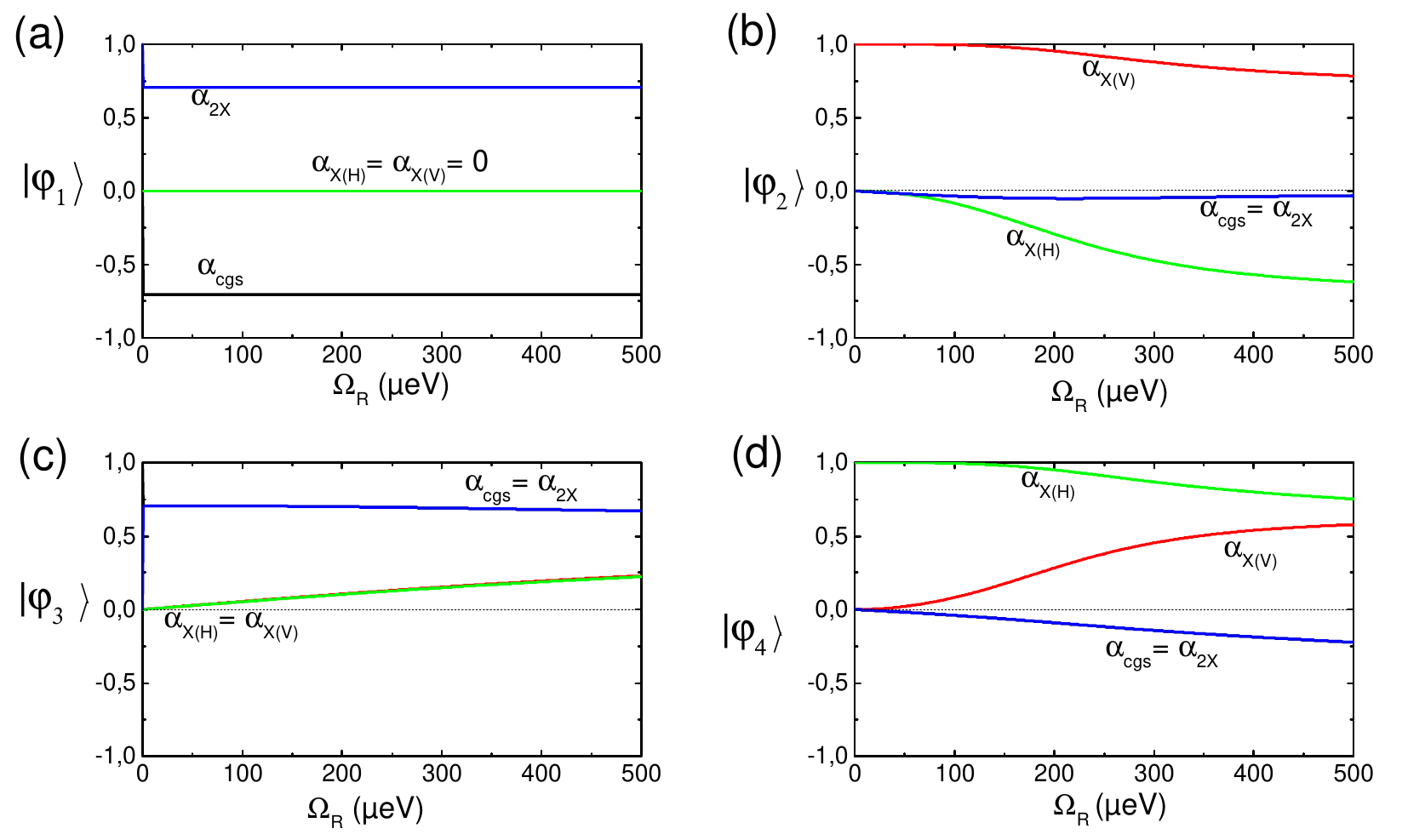}
\caption{\label{fig:Figure3S} (Color online) Coefficients $\alpha_{i}$ of the dressed states in the bare state basis as a function of the Rabi frequency $\Omega_{R}$ for the state $\ket{\varphi_{1}}$ in (a), for the state $\ket{\varphi_{2}}$ in (b), for the state $\ket{\varphi_{3}}$ in (c) and for the state $\ket{\varphi_{4}}$ in (d).}
\label{Figure3S}
\end{figure}

To analyze the evolution of the dressed state in this regime we plot in Fig.\ref{fig:Figure3S} the real parts of the normalized coefficients $\alpha_{i}$ of the eigenstates $\ket{\varphi_{1/2/3/4}} = \sum_{j} \alpha_{j} \ket{j}$ in the bare state basis with $j = cgs, X_{0}(H), X_{0}(V), 2X_{0}$ as a function of the Rabi energy $\Omega_{R}$. While the state $\varphi_{1}$ in Fig. \ref{fig:Figure3S}a with a zero eigenenergy is independent of the Rabi energy $\Omega_{R}$, the three other states $\ket{\varphi_{2/3/4}}$ in Fig.\ref{fig:Figure3S} b-d exhibit a Rabi frequency dependent mixing angle such that $\alpha_{j} = \alpha_{j} (\Omega_{R})$. We make three remarks; (i) For the three states $\ket{\varphi_{2/3/4}}$ the components $\alpha_{2X} = \alpha_{cgs}$ are equal due the resonant two-photon coupling of the $cgs$ to biexciton $2X_{0}$. (ii) The excitonic components of $\ket{\varphi_{2/3/4}}$ are unequal $\alpha_{X_{0}(H)} \neq \alpha_{X_{0}(V)}$ due to the detuning of the exciton states $X_{0}(H)$ and $X_{0}(V)$ from the excitation laser energy induced by the fine structure $\delta_{0}$ and the binding energy $\Delta_{b}$. We note here, that for the state $\ket{\varphi_{2}}$ the excitonic components $X(H)$ and $X(V)$ enter with different signs corresponding to an opposite relative phase, while for states $\ket{\varphi_{3/4}}$ both components have a positive sign. This will be essential for the polarization of the transitions between the dressed eigenstates discussed later in this supplemental material.  

\begin{figure}
\includegraphics[width=1\columnwidth]{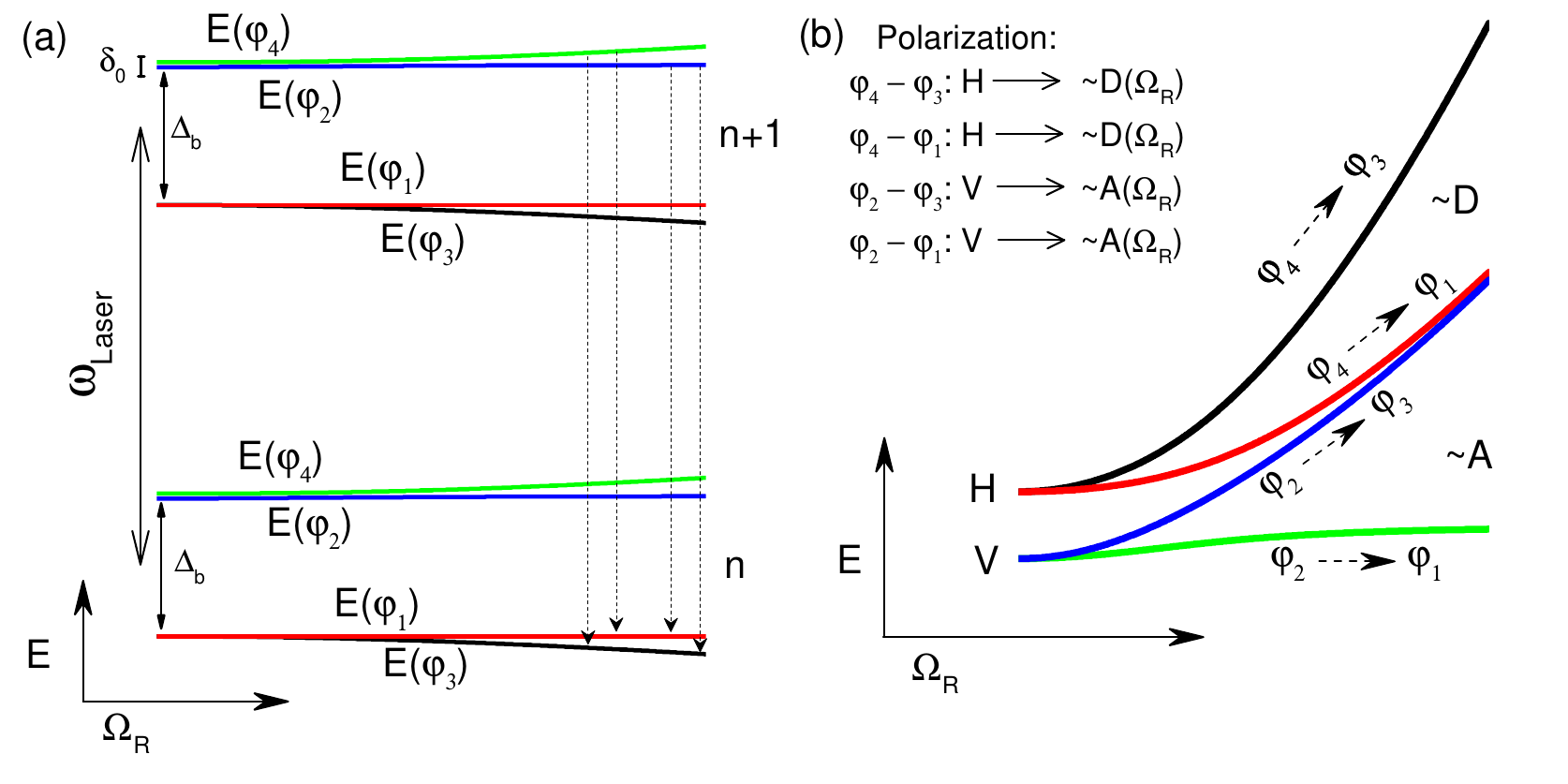}
\caption{\label{fig:Figure4S} (Color online) (a) Evolution of the eigenenergies $E_{\varphi_{i}}$ of the dressed states in a frame rotating with the laser frequency $\omega_{L}$ as a function of the Rabi energy $\Omega_{R}$. In the experiment we monitor transitions between the dressed states $\varphi_{i} \rightarrow \varphi_{j}$ of the manifolds $n$ and $n+1$ indicated by the dashed arrows. (b) Resulting energies of the dressed state transitions $\varphi_{i} \rightarrow \varphi_{j}$. For increasing Rabi energy the polarization of the transitions is rotated from $H(V)$ to $D(A)$.}
\label{Figure4S}
\end{figure}

In Fig.\ref{fig:Figure4S} we plot the eigenenergies of the dressed states $E(\varphi_{i})$ for all four states $\ket{\varphi_{i}}$ analyzed in Fig.\ref{fig:Figure3S} in a frame rotating with the laser frequency $\omega_{L}$. For QDs the binding energy $\Delta_{b}$ commonly exceeds the fine structure $\delta_{0}$ by more than one order of magnitude $\Delta_{b} >> \delta_{0}$ \cite{bayer1998exciton, bayer2002fine}. Thus, the binding energy $\Delta_{b}$ also induces a comparably larger split between the dressed eigenenergies $E_{\varphi_{2/4}}$ and $E_{\varphi_{1/3}}$ compared to the fine structure inducing a split between the dressed states energies $E_{\varphi_{4}}$ and $E_{\varphi_{2}}$. 

In the experiment, we monitor the transitions $\phi_{i} \rightarrow \phi_{j}$ between the dressed states $i$ and $j$ that are blue detuned from the excitation laser energy by approximately $\sim \Delta_{b}$ and correspond to the bare state exciton transitions $X_{0}(H/V) \rightarrow cgs$ for very low Rabi energy $\Omega_{R} << \delta_{0}, \Delta_{b}$. The four transitions are indicated by dashed arrows in Fig.\ref{Figure4S}a and the evolution of their energies is plotted in Fig.\ref{fig:Figure4S}b as a function of the Rabi energy $\Omega_{R}$. In the dressed state basis the transitions correspond to transitions between the manifolds $n+1$ and $n$ that are: (i) The two transitions $\varphi_{4}^{n+1} \rightarrow \varphi_{3}^{n}$ and $\varphi_{4}^{n+1} \rightarrow \varphi_{1}^{n}$ that emerge from the bare state transition $X_{0}(H) \rightarrow cgs$ at $\Omega_{0}$. (ii) The two transitions $\varphi_{2}^{n+1} \rightarrow \varphi_{3}^{n}$ and $\varphi_{2}^{n+1} \rightarrow \varphi_{3}^{n}$ that emerge from the bare state transition $X_{0}(H) \rightarrow cgs$ at $\Omega_{R} = 0$. Notably, the Rabi energy $\Omega_{R}$ dependence of the four transitions $\varphi_{2/4} \rightarrow \varphi_{1/3}$ in Fig. \ref{fig:Figure4S}b resembles the pattern observed in the main section of the letter.

Next we analyze the polarizations of the dressed state transitions $\varphi_{i} \rightarrow \varphi_{j}$. Here we use that the Poincar\'{e} sphere of the photonic polarizations spanned by $\ket{H}$ and $\ket{V}$ is mapped one-to-one onto the Bloch sphere of the exciton spin states spanned by $\ket{X_{0}(H)}$ and $\ket{X_{0}(V)}$ (that are created by the corresponding photons) \cite{muller2013all}. Thus we can estimate the probability $P(A/D)$ to find a photon created by spontaneous emission in a diagonally (anti-diagonally) polarized state $\ket{D/A} = \frac{1}{\sqrt{2}} (\ket{H} \pm \ket{V}$ from/to the neutral exciton $X_{0}$ using the expectation values of the corresponding operates for the optically allowed transitions. The system can decay by spontaneous emission from the biexciton to the exciton $2X_{0} \rightarrow X_{0}$ and from the exciton to ground state $X_{0} \rightarrow cgs$. Using the one-to-one mapping of the polarization sphere onto the exciton states, the sum of the normalized expectation values of the collapse operators

\begin{equation}
2X_{0} \rightarrow X_{0}(D/A)  :  \ket{X_{0}(D/A)} \bra{2X_{0}} = \frac{1}{\sqrt{2}} ( \ket{X_{0}(H)} \pm \ket{X_{0}(V)}) \bra{2X_{0}}
\label{eq:trans2X}
\end{equation}

and 

\begin{equation}
X_{0}(D/A) \rightarrow cgs  :  \ket{cgs} \bra{X_{0}(D/A)} = \frac{1}{\sqrt{2}} ( \ket{cgs} \bra{X_{0}(H)} \pm \ket{X_{0}(V)}) 
\label{eq:transX0}
\end{equation} 

correspond to the probabilities $P(D/A)$ to find a photon with polarization $D/A$ emitted from the system. We remark here that the polarizations of the two photons from the biexciton cascade $2X_{0} \rightarrow X_{0} \rightarrow cgs$ are independent (not entangled) since the fine structure $\delta_{0}$ well exceeds the linewidths of the transitions.

\begin{figure}
\includegraphics[width=1\columnwidth]{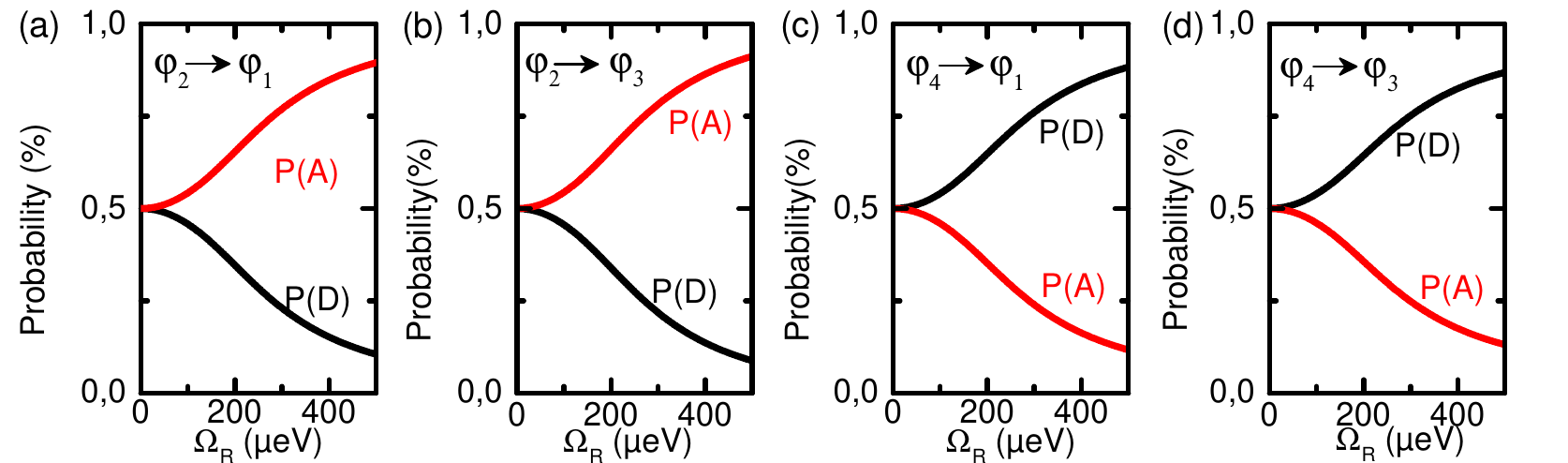}
\caption{\label{fig:Figure5S} (Color online) Relative normalized probabilities $P(A/D)$ to detect a photon with polarization $A$ and $D$ from a transition between the dressed states for $\varphi_{2} \rightarrow \varphi_{1}$ in (a), $\varphi_{2} \rightarrow \varphi_{3}$ in (b), $\varphi_{4} \rightarrow \varphi_{1}$ in (c) and $\varphi_{4} \rightarrow \varphi_{3}$ in (d).}
\label{Figure5S}
\end{figure}

We now use these collapse operators written in the diagonal (anti-diagonal) basis to analyze the polarizations of the photons emitted by spontaneous emission for transitions between the dressed states $\varphi_{i} \rightarrow \varphi_{j}$. The expectation value (degree of polarization of the transition) to find a photon with polarization $P(D/A)$ using equation \ref{eq:trans2X} and \ref{eq:transX0} then corresponds to 

\begin{equation}
 P(D/A) = \frac{ \braket{\varphi_{j}}{X_{0}(D/A)} \braket{2X_{0}}{\varphi_{i}} + \braket{\varphi_{j}}{cgs} \braket{X_{0}(D/A)}{\varphi_{i}} }{ \sum_{m=D,A} \braket{\varphi_{j}}{X_{0}(m)} \braket{2X_{0}}{\varphi_{i}} + \braket{\varphi_{j}}{cgs} \braket{X_{0}(m)}{\varphi_{i}} }  
\end{equation}.

In Fig. \ref{fig:Figure5S} a-d, we plot the normalized probabilities $P(D/A)$ for the four dressed state transitions $\varphi_{2/4} \rightarrow \varphi_{1/3}$ that we resolve experimentally in the main part of the letter. For low Rabi energy $\Omega_{R}$, all probabilities of $P_{\varphi_{i} \rightarrow \varphi_{j}}(D/A)$ converge towards $1/2$. This corresponds to the convergence of the transition polarizations towards $H$ and $V$ of the fine structure split bare state transitions $X_{0}(H/V) \rightarrow cgs$ at $\Omega_{R} = 0$. Thus for low Rabi frequencies $\Omega_{R}$, we obtain almost equal probabilities to detect a photon in $D$ or $A$. However, we observe in Fig.\ref{fig:Figure5S} a-d that for increasing Rabi energy $\Omega_{R}$ the differences between $P(D)$ and $P(A)$ increase corresponding to a rotation of the transitions polarizations. The polarizations of the two energetically lower transitions $\varphi_{2} \rightarrow \varphi_{1/3}$ rotate towards $A$ in (a) and (b) while the polarizations of the two energetically higher transitions $\varphi_{4} \rightarrow \varphi_{1/3}$ in c (d) are rotated towards $D$ in (c) and (d).

Physically we interpret this behavior by considering the evolution of the transitions eigenenergies in Fig.\ref{fig:Figure3S}b: While initially the fine structure split $\delta_{0}$ shields the transitions polarizations $H$ and $V$, it starts to be rotated into the polarization basis of the excitation laser with increasing Rabi frequency $\Omega_{R} > \delta_{0}$. For an excitation laser that is diagonally polarized along $D$ as in the experimental set up, we indicate in Fig.\ref{fig:Figure4S}b that the polarizations of the transitions $\varphi_{4}\rightarrow \varphi_{3/1}$ ($\varphi_{2}\rightarrow \varphi_{3/1}$) are increasingly polarized along $D$ (and orthogonally to it along $A$). Since we use a cross-linearly polarized excitation - detection scheme with polarizations $D - A$ to suppress the stray light of the laser, we only resolve the lower energy pair of dressed state transitions $\varphi_{2}\rightarrow \varphi_{3/1}$ for increasing Rabi frequencies $\Omega_{R}$ in the experiment.

\section{Details of the numerical calculations}
\label{sec:theory}

In order to find the steady state solutions for the reduced density matrix $\rho_{ss}$ of the Hamiltonian $H = H_{QD} + H_{int} + H_{QD-phonon}$, we perform numerical calculations of the system using the Quantum Toolbox in Python \cite{qutip, qutip2}. We include dissipation and the exciton - LA phonon interactions by an effective phonon master equation using Lindblad scattering terms following the model described in Ref. \cite{ge2013}. The master equation in the bare state basis $\ket{cgs}$, $\ket{X_{0}(H)}$, $\ket{X_{0}(V)}$ and $\ket{2X_{0}}$ then reads:

\begin{align}
\begin{split}
\frac{\partial\rho}{\partial t}&=-\frac{i}{\hbar}[H',\rho]+ \sum_{j=X_0(V),X_0(H)}(\frac{\gamma_{2X_{0}-j}}{2} L[\ket{j}\bra{2X_{0}}]+\frac{\gamma_{j-cgs}}{2} L[\ket{cgs}\bra{j}]+\frac{\gamma_{j-2X_{0}}^{ph}}{2}\\
&L[\ket{2X_{0}}\bra{j}]+\frac{\gamma_{cgs-j}^{ph}}{2} L[\ket{j}\bra{cgs}]+\frac{\gamma_{2X_{0}-j}^{ph}}{2} L[\ket{j}\bra{2X_{0}}]+\frac{\gamma_{j-cgs}^{ph}}{2} L[\ket{cgs}\bra{j}]+\\
&\Gamma_{j}^{cd}(\ket{j}\bra{cgs}\rho\ket{j}\bra{cgs}+\ket{cgs}\bra{j}\rho\ket{cgs}\bra{j})+\Gamma_{2X_{0}}^{cd}(\ket{2X_{0}}\bra{j}\rho\ket{2X}\bra{j}+\ket{j}\\
&\bra{2X_{0}}\rho\ket{j}\bra{2X_{0}}))+ \frac{\gamma_{2X_{0}-cgs}}{2}
L[\ket{cgs}\bra{2X_{0}}]+\sum_{i=X_0(V),X_0(H),2X_{0}} \frac{\Gamma_i}{2} \ket{i}\bra{i}
\end{split}
\label{eq:polaronme}
\end{align}

Here, $L[\ket{i}\bra{j}]=\ket{i}\bra{j}\rho \ket{j}\bra{i}-\ket{j}\bra{i} \rho \ket{i}\bra{j}~+~$H.c. describes the Lindblad terms with the recombination rates $\gamma_{i-j}$, the phonon assisted transition rates $\gamma_{i-j}^{ph}$ and the phonon induced cross dephasing $\Gamma_{i-j}^{cs}$. The pure dephasing for the states $\ket{i}$ is included by the rates $\Gamma_{i}$.

We furthermore include the exciton - LA phonon interaction in the polaron-modified Hamiltonian: 

\begin{align}
\begin{split}
H'&=(\Delta_b/2-\delta_{0}/2+\hbar\delta_{p})\ket{X_0(V)}\bra{X_0(V)}+(\Delta_b/2+\delta_{0}/2+\hbar\delta_{p})\ket{X_0(H)}\\
&\bra{X_0(H)}-\hbar\delta_{p} \ket{2X_{0}}\bra{2X_{0}}+\hbar/2~\Omega_V' (\ket{cgs}\bra{X_0(V)}+\ket{X_0(V)}\bra{cgs}+\ket{X_0(V)}\\
&\bra{2X_{0}}+\ket{2X_{0}}\bra{X_0(V)})+\hbar/2~\Omega_H'(\ket{cgs}\bra{X_0(H)}+\ket{X_0(H)}\bra{cgs}+\ket{X_0(H)}\\
&\bra{2X_{0}}+\ket{2X_{0}}\bra{X_0(H)})
\end{split}
\end{align}

with half of the exciton binding energy $\Delta_b$ and the anisotropic exciton fine structure splitting $\delta_{0}$ between the neutral exciton states $X_{0}$. Using equation \ref{eq:polaronme} we compute the steady state solution of the density matrix with:

\begin{equation}
\rho_{ss}=\frac{\partial \rho(t \to \infty)}{ \partial t} = -\frac{i}{\hbar}[H,\rho_{red}]+L[\rho_{red}] \to 0
\label{eq:steadystate}
\end{equation}

for a given Rabi frequency $\Omega_R$ to calculate the correlation function $G(\tau)$ for a discrete set of time delays $\tau$. We obtain the fluorescence spectrum as the Fourier transform of the correlation function $G(\tau)$ as \cite{scully1997}:

\begin{equation}
S(\omega)=\frac{1}{\pi} \operatorname{Re} \int \limits_0^\infty \! G(\tau) e^{i\omega\tau} \, d\tau
\label{eq:spec}
\end{equation}.

\begin{table}
    \caption{Parameters used in the numerical simulations taken from \cite{krummheuer2002}.}
    \begin{tabular}{| l | l |}
    \hline
    Quantity & Value \\ \hline
    Exciton radiative decay rate & $\gamma_{X_0(H/V)-cgs}=1.54$ ns$^{-1}$ \\ \hline
    Exciton dephasing rate & $\Gamma_{X_0(H/V)}=2.56$ ns$^{-1}$\\ \hline
    Biexciton radiative decay rate & $\gamma_{2X_{0}-X_0(H/V)}=1.54$ ns$^{-1}$ \\ \hline
		Biexciton dephasing rate & $\Gamma_{2X_{0}}=2.56$ ns$^{-1}$ \\ \hline
		Two photon radiative decay rate & $\gamma_{2X_{0}-cgs}=0.002$ ns$^{-1}$\\ \hline
    Biexciton binding energy & $E_B=1.86$ meV \\ \hline
		Finestructure splitting & $\delta=32$ $\mu$eV \\ \hline
		Temperature & $T=4.2$ K \\ \hline
		Density of GaAs & $\rho_{GaAs} =5.37$ g/cm$^3$ \\ \hline 
		Longitudinal sound velocity of GaAs & $v_s=5110$ m/s\\ \hline
		Electron Deformation potential & $D_e=-14.6$ eV \\ \hline
		Hole Deformation potential & $D_h=-4.8$ eV \\ \hline
		Electron/Hole confinement length & $d=8.06$ nm \\ \hline
		
    \end{tabular}
\label{tab:parameters}		
\end{table}

The QD-LA phonon interaction enters the polaron modified Hamiltonian $H'$ in two places: Firstly by a polaron shift $\delta_{p}$ of the bare state eigen-energies and secondly by a renormalization of the Rabi frequencies $\Omega_{H/V} \rightarrow \Omega_{H/V}'$. The QD - LA phonon interaction is characterized by the phonon spectral density function $J(\omega) = \sum_{q} \lambda_{q}^{i} \delta (\omega - \omega_{q})$. Following the procedure given in the supplementary of Ref. \cite{ramsay2010phonon} we assume:

\begin{equation}  
J(\omega)=\frac{(D_e-D_h)^2}{4\pi^2\rho_{GaAs} v_s^5} \omega^3 exp(\frac{-\omega^2 d^2}{2 v_s^2})
\end{equation}

for a spherical quantum dot and equal electron and hole confinement lengths $d=d_{e/h}$. $D_e$ and $D_h$ denote the electron and hole deformation potential constants, $\rho_{GaAs}$ the density of GaAs and $v_s$ the speed of sound in GaAs. We remark that we use a linear dispersion relation $\omega_q=v_s q$ for a wave vector $q$. The interaction between the quantum dot and the LA-phonon bath results in a renormalization of the Rabi frequency:

\begin{equation}  
\Omega_{V/H}'= \Omega_{V,H} exp[ \, \frac{1}{2} \int \limits_0^\infty \! J(w)/w^2 coth(\hbar\omega/2k_b T)] \, d\omega
\end{equation}

where $k_b$ is the Boltzmann constant and $T$ the sample temperature. We calculate the polaron shift as:

\begin{equation}  
\delta_p=\int \limits_0^\infty \! J(w)/w \, d\omega
\end{equation}

The phonon induced scattering rates corresponding to incoherent excitation from the state $\ket{j}$ to the state $\ket{i}$ and phonon mediated relaxation from the state $\ket{i}$ to the state $\ket{j}$ can be calculated as:

\begin{equation}
\gamma_{i-j}^{ph}= \Omega_{H/V}'^{2}/2 \operatorname{Re}[ \int \limits_0^\infty \! exp(\pm \Delta \omega_{i-j} \cdot \tau)(exp(\Phi(\tau))-1) \, d\tau ].
\end{equation}

The phonon induced cross dephasing is described by:

\begin{equation}
\Gamma_{i}^{cd}= \Omega_{H/V}'/2 \operatorname{Re} \int \limits_0^\infty \! cos(\Delta \omega_{i-j} \cdot \tau)(1-exp(-\Phi(\tau))) \, d\tau
\end{equation}

with 

\begin{equation}  
\Phi(\tau)=\int \limits_0^\infty \! J(w)/w^2 [coth(\hbar\omega/2k_b T) cos(\omega\tau)-i sin(\omega\tau)] \, d\omega
\end{equation}

and the frequency detuning $\Delta \omega_{i-j}=\omega_L-\omega_{i-j}$ of the excitation field $\omega_L$ from the transition $i \rightarrow j$. Finally we present the parameters, that we use for the numerical simulations in table \ref{tab:parameters}.

\section{Phonon assisted pump rates}
\label{sec:phonon}

In the previous section \ref{sec:theory} we obtain incoherent phonon-assisted pump rates that are on the order of the radiative decay rate $\gamma^{ph}(P_{exc}) > \gamma_{i \rightarrow j} $ in agreement with the population distribution measured in the letter. Since up to date only significantly smaller phonon assisted excitation rates are reported for cw excitation \cite{weiler2012, ge2013}, we demonstrate in this section experimentally that the incoherent phonon assisted pump $\gamma^{ph}$ rates can exceed the radiative recombination rates $\gamma_{X_{0} \rightarrow cgs}$ of the neutral exciton state in a QD.  

\begin{figure}
\includegraphics[width=1\columnwidth]{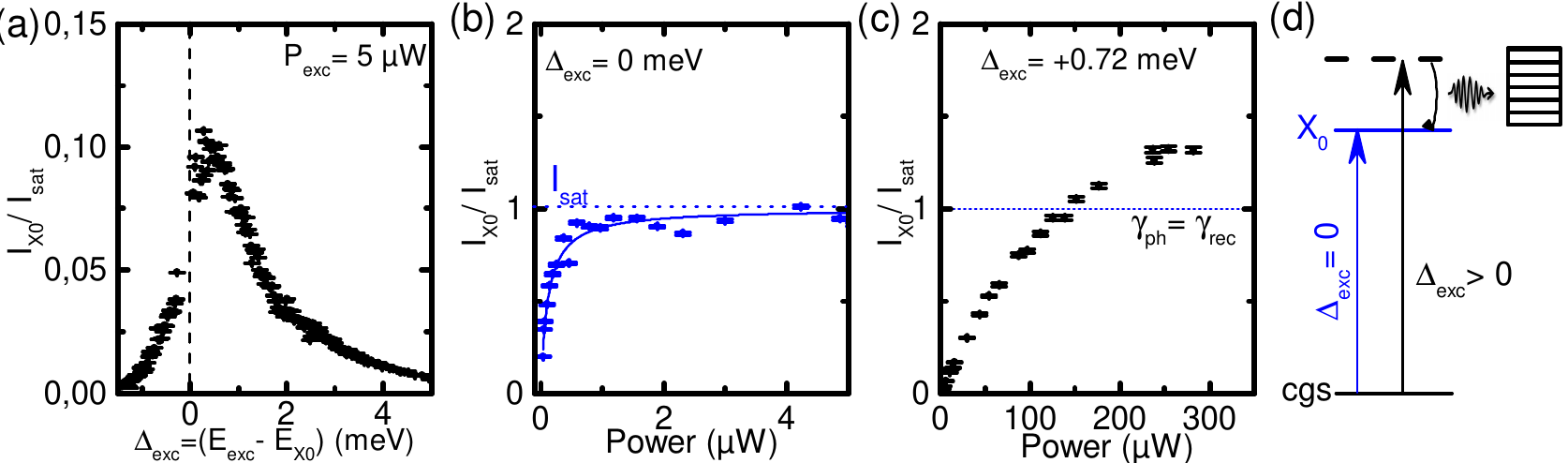}
\caption{\label{fig:Figure2S} (Color online) (a) Luminescence of the $X_{0} \rightarrow cgs$ transition as a function of the excitation detuning $\Delta_{exc}$. Emission under blue detuned excitation $\delta_{exc}> 0 \, \milli \electronvolt$ corresponds to the phonon assisted excitation process of $X_{0}$ as depicted in (d). (b) Emission intensity of the neutral exciton transition $I_{X0}$ for resonant excitation $\Delta_{exc} = 0 \, \milli \electronvolt$ with a fit indicated by the black line that converges towards the incoherent limit. (c) Emission intensity of the neutral exciton $I_{X0}$ for blue detuned excitation in the phonon sideband. (d) Schematic illustration of the excitation process: (i) Resonant excitation for $\Delta_{exc} = 0 \, \milli \electronvolt$ of the $cgs \rightarrow X_{0}$ transition (black) and (ii) blue detuned excitation that leads to a phonon assisted population transfer into the state $X_{0}$ (blue).}
\label{Figure2S}
\end{figure}

In Fig. \ref{fig:Figure2S}b we present the well-known saturation behavior of the luminescence (proportional to the exciton population $n_{X_{0}}$) $I_{X_{0}} \propto n_{X_{0}}$ of the neutral exciton $X_{0}$ of a QD under resonant excitation $\Delta_{exc} = 0 \, \milli \electronvolt$ as a function of the excitation power $P_{exc}$ \cite{muller2007}. The population saturates against the incoherent limit $n_{X_{0}} = \frac{1}{2}$. To investigate the phonon assisted excitation process that is schematically illustrated in Fig. \ref{fig:Figure5S}d, we perform measurements in dependence of the excitation detuning $\Delta_{exc}$ on the neutral exciton transition $cgs \rightarrow X_{0}$. The results are presented in Fig. \ref{fig:Figure2S}a. For a blue detuning $\Delta_{exc} > 0 \, \milli \electronvolt$, we observe luminescence from the neutral exciton $X_{0} \rightarrow cgs$ transition. However, it is strongly suppressed for a negative excitation detuning $\Delta_{exc} < 0$. The detuning dependent shape of the phonon assisted excitation process results from the asymmetric LA - phonon QD interaction spectrum $J(\Delta \omega_{exc})$ discussed in the previous section and Ref. \cite{glassl2013, Ardelt2014, quilter2015, weiler2012, bounouar2015}. Note, that for the two-photon excitation scheme presented in the letter, we are blue detuned from the $X_{0} \rightarrow 2X_{0}$ transition by $\Delta_{b} = 0.92 \, \milli \electronvolt$ leading to an effective LA - phonon interaction. To demonstrate that the incoherent phonon assisted excitation rate $\gamma^{ph}$ can exceed the radiative recombination rate of the QD, we set the energy of the excitation laser to a detuning of $\Delta_{exc} = +0.72 \, \milli \electronvolt$. The power dependent luminescence $I_{X_{0}}$ is normalized to the maximum of the luminescence for resonant excitation of the $X_{0}$ (Fig. \ref{fig:Figure2S}b) and presented in Fig. \ref{fig:Figure2S}c. For excitation powers $P_{exc} > 150 \, \micro \watt$, the luminescence exceeds the maximum luminescence $I_{X_{0}} = 1$ corresponding to the case $\gamma^{ph} > \gamma_{X_{0} \rightarrow cgs}$. This also corresponds to a population inversion $n_{X_{0}} > n_{cgs}$ of the two level system illustrated in Fig. \ref{fig:Figure2S}d that consists of the ground state $cgs$ and the neutral exciton $X_{0}$ state. We conclude that including the QD - LA phonon interaction in the theoretical model is essential to obtain an accurate population distribution in the four level system as discussed in the letter.    

\newpage
\bibliography{Papers}

\end{document}